\documentclass[conference,a4paper]{IEEEtran}

\usepackage[utf8]{inputenc} 
\usepackage[T1]{fontenc}
\usepackage{url}              
\usepackage{cite}             

\usepackage[cmex10]{amsmath}  
\usepackage{mleftright}       
\mleftright                   

\usepackage{graphicx}         
\usepackage{booktabs}         

\usepackage{bm,dsfont}
\usepackage{soul}
\usepackage{amssymb}
\newtheorem{thm}{Theorem}[section]

\newtheorem{lem}{Lemma}
\newtheorem{defn}{Definition}

\usepackage{xcolor}
\newcommand{\E}{\mathds{E}}

\newcommand{\dtv}{d_\mathrm{TV}}
\newcommand{\cb}{\mathcal{C}^{(n)}}

\begin{document}

\title{Conditional Rate-Distortion-Perception Trade-Off} 


%
%

\author{%
  \IEEEauthorblockN{Xueyan Niu\IEEEauthorrefmark{1},
                    Deniz G\"und\"uz\IEEEauthorrefmark{2}\IEEEauthorrefmark{1},
                    Bo Bai\IEEEauthorrefmark{1},
                    and Wei Han\IEEEauthorrefmark{1}}
  \IEEEauthorblockA{\IEEEauthorrefmark{1}%
                    Theory Lab, 2012 Labs, Huawei Technologies Co. Ltd.,
                    \{niuxueyan3, baibo8, harvey.hanwei\}@huawei.com}
  \IEEEauthorblockA{\IEEEauthorrefmark{2}%
                    Department of Electrical and Electronic Engineering, Imperial College London, London, UK,
                    d.gunduz@imperial.ac.uk}
}

\maketitle

\begin{abstract}
  Recent advances in machine learning-aided lossy compression are incorporating perceptual fidelity into the rate-distortion theory. 
  In this paper, we study the rate-distortion-perception trade-off when the perceptual quality is measured by the total variation distance between the empirical and product distributions of the discrete memoryless source and its reconstruction. We consider the general setting, where two types of resources are available at both the encoder and decoder: a common side information sequence, correlated with the source sequence, and common randomness. We consider both the strong perceptual constraint and the weaker empirical perceptual constraint. The required communication rate for achieving the distortion and empirical perceptual constraint is the minimum conditional mutual information, and similar result holds for strong perceptual constraint when sufficient common randomness is provided and the output along with the side information is constraint to an independent and identically distributed sequence. 
\end{abstract}

\section{Introduction} \label{sec:intro}
The practice of lossy compression is rich with recent success stories,  where machine learning driven methods outperform traditional codecs in image, video, and audio compression. In data-driven approaches, training often relies on minimizing discrepancies between the source and reconstruction distributions (e.g., cross-entropy, Wasserstein distance) to capture the quality of the reconstructed data perceived by humans \cite{santurkar2018generative}. Given a source distribution and a distortion measure, the rate-distortion theory studies the minimum rate required to achieve the target distortion level measured by an additive distortion measure, and it has been the theoretical framework for the design and evaluation of lossy compression codecs.
However, it has been shown that the mean-squared error does not reflect the perceptual quality of reconstructions \cite{blau2018perception,dahl2017pixel,tschannen2018deep}.
In \cite{blau2019rethinking}, Blau and Michaeli propose the information rate-distortion-perception function to characterize the three-way tradeoff between the rate, distortion, and perceptual quality by imposing a constraint on the distribution of the reconstruction.

The additional perceptual constraint is intimately related to the theory of coordination. In \cite{cuff2010coordination}, two notions of coordination, \textit{empirical coordination} and \textit{strong coordination}, are distinguished. Both require the output distribution to be close to a target distribution in terms of total variation. As the names suggest, empirical coordination concerns the empirical distribution (type), while strong coordination deals with the joint distribution over the block of symbols. In this paper, we also study two notions of perceptual quality corresponding to the empirical and strong coordination requirements.  In machine learning scenarios, the empirical perceptual constraint can be interpreted using generative modeling, such as generative adversarial networks (GANs) \cite{goodfellow2020generative}, where the goal of the generator is to minimize a certain divergence between the data distribution and the empirical distribution of the synthetic samples. The typicality of empirical distributions has been studied by \cite{raginsky2012empirical} and \cite{kramer2007communicating} in different contexts. The strong perceptual constraint demands that the order of the samples is also preserved, which is related to natural language processing problems \cite{manning1999foundations}, where grammar and word order are essential in the reconstruction.

Unlike the \textit{average} distortion measures, the perceptual fidelity compares two probability distributions. 
One of the insights developed in \cite{matsumoto2018introducing} is that the $n$-letter perception fidelity evaluated by total variation between product distributions precludes single-letterization of the rate region, and that a channel resolvability code may be necessary. Accordingly, much work has focused on the so-called \textit{perfect realism} condition, which requires the coding to be distribution-preserving, i.e., the distribution of the output approximates arbitrarily well the source distribution in total variation. Saldi, Linder, and Y{\"u}ksel \cite{saldi2015output} connect distribution-constrained lossy coding with distributed channel synthesis. In their setup, apart from the distortion constraint, the output sequence is restricted to follow a given distribution of an independent and identically distributed (i.i.d.) sequence. Thus, the \textit{perfect realism} setting in \cite{wagner2022rate} can be viewed as a special case of \cite{saldi2015output}. 
The output distribution constraint requires vanishing total variation error, whereas in our setting, the perceptual constraint allows a bounded total variation. The extension is not straightforward, as in the achievability proof, an idealized distribution which approximates the source-reproduction joint distribution is used to facilitate the analysis of the distortion and perception performances. Another challenge arises from the common side information, for which we tailor the local channel synthesis argument \cite{cuff2013distributed} to our setup.

Recent research suggests that common randomness plays an important role in achieving the rate-distortion-perception function \cite{theis2021advantages,theis2021coding,wagner2022rate,chen2022rate}. As noted by these works, the common randomness, though being a scarce resource, can be realized by agreeing upon some seeds for the pseudo-random number generator in advance, and the seeds can also be encoded as messages. 
Nevertheless, as shown in \cite{cuff2010coordination,chen2022rate}, in the setting of empirical coordination, common randomness is not necessary.
We consider a general situation where the sender and the receiver possess two types of resources: common side information and common randomness. The common side information is a random process that may be correlated with the source, providing additional information for the reconstruction; while the common randomness is a random variable observed by both the sender and the receiver, independent of the source signal, and it is usually assumed to be uniformly distributed on a finite set.  We present rate regions when the amount of common randomness is constrained. The conditional rate-distortion-perception function is derived for empirical perceptual constraint, and for the strong perceptual constraint when sufficient common randomness is available and the output along with the side information is limited to an i.i.d. sequence.

\section{Problem Setup and Main Result}\label{sec:setup}
 
Let $\{X_i\}_{i=1}^\infty$, $\{Z_i\}_{i=1}^\infty$ be memoryless sources drawn from finite alphabets $\mathcal{X}$ and $\mathcal{Z}$ according to a joint distribution $P_{XZ}(x,z).$ We drop the arguments of the distribution when it does not cause ambiguity. The $n$-sequence $(X_1,X_2,\ldots,X_n)$ is denoted by $X^n,$ and $(X_t,X_{t+1},\ldots,X_n)$ is denoted by $X_{t}^n.$ So $P_{X^nZ^n}=\prod_{i=1}^n P_{XZ}.$ 
We use $[m]$ to denote the set $\{1,2,\ldots, \lfloor m \rfloor\}$ for $m>0,$ and $\mathcal{C}$ to denote the codebook.
For sequences $(x^n, y^n)\in\mathcal{X}^n\times \mathcal{Y}^n,$ the empirical distribution is defined as
\begin{equation}\label{eq:emp}
\hat{P}_{x^n, y^n}(x,y) = \frac{1}{n}\sum_{i=1}^n \mathbf{1}\{(x_i,y_i)=(x,y)\}.
\end{equation}
Extensions to multiple arguments can be defined similarly.

\begin{figure}[tbp]
  \centering
  \includegraphics[width=0.35\textwidth]{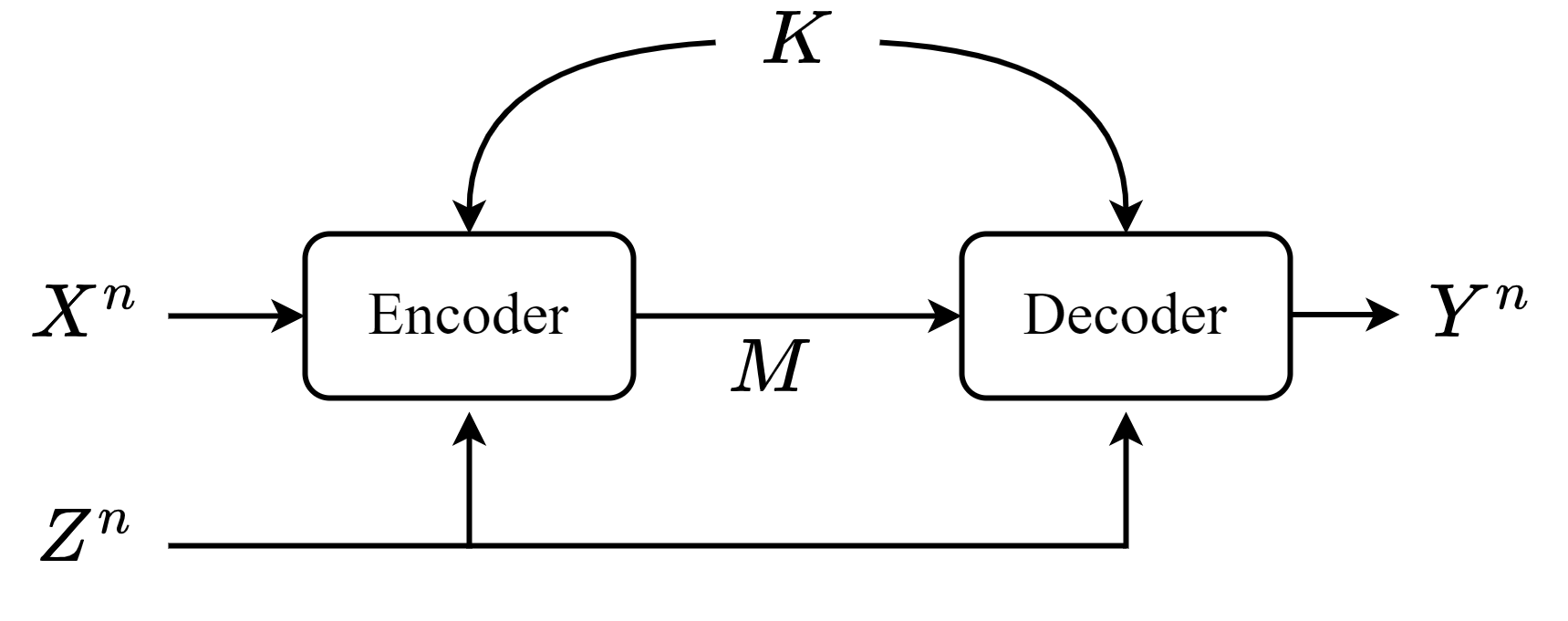}
  \caption{Rate-distortion-perception model with common randomness and common side information.}
  \label{fig:model}
  \vskip -0.1in
\end{figure}

Consider the model depicted in Fig. \ref{fig:model}, in which the encoder and the decoder have access to a shared source of randomness $K$ uniformly distributed over $[2^{nR_0}].$ The encoder observes source $X^n$ and side information $Z^n,$ and selects a message $M$. The decoder observes the message, and it also has access to the common randomness $K$ and side information $Z^n,$ and tries to recover input $X^n.$ 
\begin{defn}
    An $(n, 2^{nR}, 2^{nR_0})$ code with common randomness and common side information consists of an encoding function
    \[
    f_n: \mathcal{X}^n \times \mathcal{Z}^n \times [2^{nR_0}] \mapsto [2^{nR}] \quad \text{(possibly stochastic)}
    \]
    and a decoding function
    \[
    g_n: [2^{nR}]\times \mathcal{Z}^n \times [2^{nR_0}] \mapsto \mathcal{X}^n \quad \text{(possibly stochastic)}.
    \]
\end{defn}
As usual, the degree of distortion between the source sequence and the reconstruction is the average of a per-letter distance. 
\begin{defn}\label{defn:distortion}
    Given a per-letter distortion measure
    $D:\mathcal{X}\times \mathcal{X}\mapsto [0,d_{max}]$ with $d_{max}<\infty,$
    the (average) distortion between two sequences $x^n$ and $y^n$ is defined as
    \[
    D(x^n, y^n) := \frac{1}{n}\sum_{i=1}^n D(x_i, y_i).
    \]
\end{defn}
Much work has been focused on total variation distance as a measure of coordination \cite{cuff2010coordination,raginsky2012empirical,cuff2013distributed,yassaee2014achievability,matsumoto2018introducing,wagner2022rate}, and we also use total variation to evaluate the perceptual quality. 
\begin{defn}
    The total variation between two distributions $P_X$ and $P_Y$ defined on the same $\sigma$-algebra $(\mathcal{X},\mathcal{F})$ is
    \[
    \dtv (P_{X}, P_{Y}) = \sup_{A\subseteq \mathcal{X}} |P_{X}(A) - P_{Y}(A)|.
    \]
\end{defn}
Inspired by the theory of coordination \cite{cuff2010coordination}, we define two notions of achievability with respect to empirical and strong perceptual constraints. In the case of the empirical one, permutation of the $n$-sequence does not affect the perception, whereas for the strong perceptual constraint, the order of the pairs $(X_i, Y_i)$ matters.
\begin{defn}[Achievability]
The tuple $(R, R_0, \Delta, \Pi)$ is achievable with respect to empirical and strong perceptual constraints if for any $\epsilon>0,$ there exists a sequence of $(n, 2^{n(R+\epsilon)}, 2^{n(R_0+\epsilon)})$ codes $(f_n, g_n)$ such that
\begin{align*}
    \E_{P} [D(X^n, Y^n)] &\leq \Delta + \epsilon, 
\end{align*}
where $Y^n = g_n(f_n(X^n,Z^n,K),Z^n,K),$
and one of the following corresponding constraints hold:
\begin{align*}
    \E_{\cb} [\dtv (\hat{P}_{X^n}, \hat{P}_{Y^n})] &\leq \Pi +\epsilon \qquad \text{(empirical perception)}\\
    \dtv ({P}_{X^n}, {P}_{Y^n}) &\leq \Pi +\epsilon. \qquad \text{(strong perception)}
\end{align*}
\end{defn}
Note that for empirical perception, the total variation is between two empirical distributions as defined in Eq \eqref{eq:emp}, while for strong perception, the total variation is between the entire coding block. In fact, as can be seen from \ref{prop:marginal} in Sec.~\ref{sec:prelim}, we have
\begin{align*}\label{eq:n-tv-bound}
    \E_{\cb} &[\dtv (\hat{P}_{X^n}, \hat{P}_{Y^n})]\leq \max_i \dtv(P_{X_i}, P_{Y_i}) 
    \leq \dtv (P_{X^n}, P_{Y^n}).
\end{align*}
We define the following rate regions.
\begin{defn}
     For arbitrarily small $\gamma>0,$ the $\gamma$-rate-distortion-perception regions with common randomness and common side information are
   \begin{align*}
        \mathcal{R}_\gamma^{(s)-} = \{ (R, R_0, \Delta, \Pi):\ &\exists\ (U,Y)\ \text{s.t.} \ \Pi<f(\Delta), R_0>0\\
        X & \perp Y\ |\ U,Z \\
        R & \geq I(X;U|Z) + \gamma\\
        R + R_0 & \geq I(Y;U|Z) + \gamma\\
        \Delta & \geq \E_{P} [D(X, Y)]\\
        \Pi & \geq \limsup_{n\rightarrow\infty}\dtv (\textstyle\prod_{i=1}^nP_{X}, \textstyle\prod_{i=1}^nP_{Y}) \}\\
        \mathcal{R}_\gamma^{(s)+} = \{ (R, R_0, \Delta, \Pi):\ &\exists\ Y \ \text{s.t.} \ \Pi\geq f(\Delta), R_0\geq 0\\
        R & \geq I(X;Y|Z) + \gamma\\
        \Delta & \geq \E_{P} [D(X, Y)]\}
    \end{align*}
    for strong perception and
        \begin{align*}
        \mathcal{R}_\gamma^{(e)} = \{ (R, R_0, \Delta, \Pi):\ &\exists\ Y \  \text{s.t.} \ R_0\geq 0\\
        R & \geq I(X;Y|Z) + \gamma\\
        \Delta & \geq \E_{P} [D(X, Y)]\\
        \Pi & \geq \dtv (P_{X}, P_{Y})\}
    \end{align*}
   for empirical perception, where $f(\Delta):=g(R(\Delta), \Delta)$ with
   \begin{align*}
       g(R, \Delta) &:=\!\!\!\!\!\!\!\!\! \inf_{\substack{P_{Y|XZ}: I(X;Y|Z)\leq R, \\ \E_{P} [D(X, Y)]\leq \Delta}} \limsup_{n\rightarrow\infty}\dtv (\textstyle\prod_{i=1}^nP_{X}, \textstyle\prod_{i=1}^nP_{Y} ) \\
       R(\Delta) &:= \inf_{P_{Y|XZ}: \E_{P} [D(X, Y)]\leq \Delta} I(X;Y|Z).
   \end{align*}
\end{defn}
\begin{defn}
    The information rate-distortion-perception regions with common randomness and side information are
    \begin{align*}
        \mathcal{R}^{(e)} = \bigcup_{\gamma>0}\mathcal{R}_\gamma^{(e)} \qquad \text{and} \qquad \mathcal{R}^{(s)} = \bigcup_{\gamma>0} (\mathcal{R}_\gamma^{(s)-} \cup \mathcal{R}_\gamma^{(s)+} ).
    \end{align*}
\end{defn}
We will prove the following theorems in Sec.~\ref{sec:achievability} and Sec.~\ref{sec:converse}.
\begin{thm}\label{thm:achievable}
For arbitrarily small $\gamma>0$ and bounded distortion measure, the rate-distortion-perception tuple $(R, R_0, \Delta, \Pi)\in \mathcal{R}_\gamma^{(s)}$ (resp. $\mathcal{R}_\gamma^{(e)}$) is achievable with respect to strong (resp. empirical) perceptual constraint.
\end{thm}
\begin{thm}\label{thm:converse}
If $(R, R_0, \Delta, \Pi)$ is achievable with respect to empirical perception, then$(R, R_0, \Delta, \Pi)\in\mathrm{cl}(\mathcal{R}^{(e)}).$
If $(R, R_0, \Delta, \Pi)$ is achievable with respect to strong  perception such that $Y^n = g_n(f_n(X^n,Z^n,K),Z^n,K)$ and $Z^n$ are jointly i.i.d. sequences, then $(R, R_0, \Delta, \Pi)\in \mathrm{cl}(\mathcal{R}^{(s)}).$
\end{thm}

The above theorems establish the rate regions for empirical perceptual constraint, and strong perceptual constraint under an i.i.d. output assumption. It is implied that common randomness $R_0$ is not necessary for empirical perceptual constraint, which can be justified using the same line of argument as in Theorem 2 of \cite{cuff2010coordination}. 
Also, in the rate region $\mathcal{R}^{(s)-},$ it is required that the strong perceptual loss $\Pi<f(\Delta).$
This is because when $\Pi\geq f(\Delta),$ the perceptual constraint is relaxed, so the problem reduces to the conventional rate-distortion problem where the common randomness is not necessary and the rate region is $\mathcal{R}^{(s)+}.$ The following lemma shows the relation between the rate regions.
\begin{lem}
$\mathcal{R}^{(s)}\subseteq \mathcal{R}^{(e)}.$
\end{lem}

Specifically, as a result of Theorem~\ref{thm:achievable} and Theorem~\ref{thm:converse}, the rate region of the empirical perceptual constraint induces a conditional rate-distortion-perception function.
\begin{thm}
    The rate-distortion-perception function with respect to empirical perceptual constraint is
    \[
    R^{(e)}(\Delta, \Pi) = \inf_{\substack{P_{Y|XZ}: \E_{P} [D(X, Y)]\leq \Delta\\ \phantom{P_{Y|XZ}:}\dtv (P_{X}, P_{Y})\leq \Pi}} I(X;Y|Z).
    \]
\end{thm}
When provided with sufficient common randomness, similar rate-distortion-perception function for strong perceptual constraint can be obtained.
\begin{thm}
    When $R_0 = \infty,$ the rate-distortion-perception function with strong perceptual constraint has the following bound
    \begin{equation*}\label{eq:s-rdp}
    R^{(s)}(\Delta, \Pi) \leq \!\!\!\!\!\!\!\!\!\!\!\! \inf_{\substack{P_{Y|XZ}:\E_{P} [D(X, Y)]\leq \Delta\\ \limsup_{n\rightarrow\infty}\dtv (\textstyle\prod_{i=1}^nP_{X}, \prod_{i=1}^nP_{Y} )\leq \Pi}} \!\!\!\!\!\!\!\! I(X;Y|Z),
    \end{equation*}
    and the bound is tight when the output sequence $Y^n = g_n(f_n(X^n,Z^n,K),Z^n,K)$ and the side information $Z^n$ are jointly i.i.d..
\end{thm}

We remark that a similar result for strong perception is obtained in \cite{chen2022rate} in the absence of correlated side information, assuming that the perception measure is tensorizable, which implicitly results in the optimality of an i.i.d. output sequence. However, since the total variation distance does not tensorize, we had to impose the i.i.d. output sequence requirement explicitly. 

\section{Preliminary Results}\label{sec:prelim}
The total variation measure has the following properties that will be useful in the proofs. 
These properties can be found in, for example, \cite{schieler2014rate}. 
Let $P,Q,R$ be probability measures on the same $\sigma$-algebra $(\mathcal{X}^2, \mathcal{F}).$
\begin{align}
\dtv (P_XP_{Y|X}, Q_XP_{Y|X}) = \dtv (P_X, Q_X). \label{prop:conditional}\tag{Property 1} \\
\dtv (P_X, Q_X) \leq \dtv (P_{XY}, Q_{XY}).
\label{prop:marginal}\tag{Property 2} \\
\dtv (P,Q) \leq \dtv (P, R) + \dtv (R, Q).
\label{prop:triangle}\tag{Property 3}
\end{align}
The total variation is convex; that is, for $\lambda_1,\lambda_2,\ldots,\lambda_n\in \mathbb{R},$
\begin{equation}
\dtv (P, \sum_{i=1}^n\lambda_i Q_i) \leq \sum_{i=1}^n \lambda_i\dtv(P, Q_i).\label{prop:convex}\tag{Property 4}
\end{equation}

The proof of the achievability relies on the following local channel synthesis result given by Corollary VII.6 of \cite{cuff2013distributed}.
\begin{lem}[\cite{cuff2013distributed}, Corollary VII.6]\label{lem:csc}
Given joint distribution $P_{UVW}$, and let $\cb=\{u^n(w^n, j)\sim \prod_{i=1}^n P_{U|W}(u_i|w_i)\}$ be a randomly generated collection of channel inputs indexed by $j\in [ 2^{nR} ]$ for any $w^n\in \mathcal{W}^n.$ Denote by $\bm Q_{V^n}$ the output distribution of applying a uniformly randomly selected codeword $U^n(w^n, J)$ through the memoryless channel $P_{V|UW}$. 
Let $\widetilde{P}$ denote the joint distribution $P_{UV|W}\hat{P}_{w^n}.$
Given an arbitrary small $\gamma>0,$ if 
$R \geq I_{\widetilde{P}}(U;V|W)+\gamma$, then
\[
\lim_{n\rightarrow\infty} \E_{\cb} [\dtv (P_{V^n|W^n=w^n}, \bm Q_{V^n|W^n=w^n}) ] = 0.
\]
\end{lem}

\section{Achievability}\label{sec:achievability}
Next, we give the proof of Theorem \ref{thm:achievable}. We focus on the strong perceptual constraint, as the line of argument carries over to empirical perceptual constraint easily.
\begin{IEEEproof}
Given $(R, R_0, \Delta, \Pi)\in \mathcal{R}^{(s)}$ and an arbitrary small $\gamma >0,$ for any $\epsilon >0,$ we can find joint distribution $\bar{P}_{XYZU}$ such that
\begin{align*}
    (X & \perp Y\ |\ U,Z)_{\bar{P}} \\
    R + \epsilon & \geq I_{\bar{P}}(X;U|Z) + \gamma\\
    R + R_0 + \epsilon  & \geq I_{\bar{P}}(Y;U|Z) + \gamma\\
    \Delta +\epsilon & \geq \E_{\bar{P}} [D(X, Y)] \\
    \Pi +\epsilon & \geq \dtv (\bar{P}_{X^n}, \bar{P}_{Y^n})
\end{align*}
Note that we can assume that the inputs follow the same process, i.e.,
$
\bar{P}_{XZ} = P_{XZ}.
$
We denote the product distribution by 
$
\bar{P}_{X^nY^nZ^nU^n}(x^n,y^n,z^n,u^n) := \prod_{i=1}^n \bar{P}_{XYZU}(x_i,y_i,z_i,u_i).
$
We construct a random codebook
\[
\cb = \{ u^n(z^n, m, m_0) \sim \prod_{i=1}^n \bar{P}_{U|Z}(u_i|z_i) \}
\]
for $z^n\in \mathcal{Z}^n$ and $(m,m_0)\in [2^{nR}]\times [2^{nR_0}].$
We take advantage of the following likelihood encoder.
\paragraph*{Encoder}
Given source sequence $(x^n, z^n)$ and a realization $m_0\in [2^{nR_0}]$ of common randomness, the encoder selects a message $m\in [2^{nR}]$ with probability proportional to $\bar{P}_{X^n|Z^nU^n}(x^n|z^n, u^n(z^n, m, m_0)),$ i.e., the behavior of the encoder can be represented by a distribution
\begin{align*}
\bm F_{M|X^nZ^nK}&(m|x^n,z^n,m_0)\\
&= \frac{\bar{P}_{X^n|Z^nU^n}(x^n|z^n, u^n(z^n, m, m_0))}{\sum_{m'\in [2^{nR}]}\bar{P}_{X^n|Z^nU^n}(x^n|z^n, u^n(z^n, m', m_0))},
\end{align*}
where $u^n(z^n, m, m_0)$ are codewords specified by the codebook $\cb.$

Let the common randomness be represented by a random variable $K$ that is uniformly distributed over $[2^{nR_0}]$ and independent of $X$ and $Z.$ Also, let the encoded message be denoted as random variable $M.$

\paragraph*{Decoder}
The decoder receives message $m,$ and it also has access to $z^n$ and common randomness $m_0.$ It then generates $y^n$ according to the distribution 
\begin{align*}
\bm G_{Y^n|Z^nMK}&( y^n\ |\ z^n, m, m_0)\\
&= \bar{P}_{Y^n|Z^nU^n}(y^n|z^n, u^n(z^n, m, m_0))
\end{align*}

The induced joint distribution according to the encoder and the decoder is 
\begin{align*}
\bm P_{X^nY^nZ^nMK} = \frac{1}{\lfloor 2^{nR_0}\rfloor}P_{X^nZ^n}\bm F_{M|X^nZ^nK}\bm G_{Y^n|Z^nMK}
\end{align*}
The distribution is stochastic because the codebook is random.

\paragraph*{Analysis}
We will consider an auxiliary distribution $\bm Q,$ such that $\bm Q$ approximates both $\bm P$ and $\bar{P}.$
We construct $\bm Q$ using the same codebook $\cb,$ such that 
\[
\bm Q_{MK}(m,m_0)=\frac{1}{\lfloor 2^{n(R+R_0)}\rfloor},\ \forall (m,m_0)\in [2^{nR}]\times [2^{nR_0}]
\]
\begin{align*}
\text{and} \qquad
    &\bm Q_{X^nY^nZ^nMK} (x^n,y^n,z^n,m,m_0)\\
    =\ & Q_{MK} (m,m_0) P_{Z^n}(z^n) Q_{U^n|Z^nMK}(u^n|z^n,m,m_0) \\
    &\bar{P}_{X^n|Z^nU^n}(x^n|z^n, u^n(z^n,m,m_0)) \\
    &\bar{P}_{Y^n|Z^nU^n}(y^n|z^n, u^n(z^n,m,m_0)) \\
    =\ & \frac{1}{\lfloor 2^{n(R+R_0)}\rfloor}
    P_{Z^n}(z^n) \mathbf{1}\{u^n = U^n(z^n,m,m_0)\} \\
    &\prod_{i=1}^n \bar{P}_{X|ZU}(x_i|z_i, u_i) \prod_{i=1}^n \bar{P}_{Y|ZU}(y_i|z_i, u_i)
\end{align*}
where
$
U^n(z^n, m, m_0) \sim \prod_{i=1}^n \bar{P}_{U|Z}(u_i|z_i).
$
Note that $\bar{P}_{X^nY^n|Z^nU^n}=\bar{P}_{X^n|Z^nU^n}\bar{P}_{Y^n|Z^nU^n}$ according to the assumption $(X\perp Y | U,Z)_{\bar{P}}$.
We make the following key observations: 
\begin{align*}
    \bm Q_{M|X^nZ^nK} &= \bm F_{M|X^nZ^nK}, \\
    \bm Q_{Y^n|Z^nMK} &= \bm G_{Y^n|Z^nMK}.
\end{align*}
The distribution $\bm Q$ matches the system encoder and decoder. 
For any $z^n\in \mathcal{Z}^n,$ let
$\widetilde{P} = \bar{P}_{XYU|Z}\hat{P}_{z^n}.$
By continuity of the mutual information, $I_{\widetilde{P}}(X;U|Z)\rightarrow I_{\bar{P}}(X;U|Z)$ and $I_{\widetilde{P}}(Y;U|Z)\rightarrow I_{\bar{P}}(Y;U|Z)$ as $n\rightarrow \infty.$
Applying Lemma \ref{lem:csc}, when $R+R_0 + 2\epsilon \geq I_{\bar{P}}(Y;U|Z)+\gamma +\epsilon,$ we have
\[
\E_{\cb} [\dtv (\bar{P}_{Y^n|Z^n=z^n}, \bm Q_{Y^n|Z^n=z^n}) ] < \epsilon_n.
\]
Since $z^n$ are i.i.d. according to $\bar{P}_{Z}=\bm Q_Z,$ we have 
\begin{align}\label{eq:tv-barpq}
    \E_{\cb} [\dtv (\bar{P}_{Y^nZ^n}, \bm Q_{Y^nZ^n}) ] < \epsilon_n.
\end{align}
For any fixed $m'_0,$ we consider the sub-codebook $\{u^n(z^n,m,m'_0)\}$ and recall the local channel synthesis lemma. 
Applying Lemma \ref{lem:csc} (cf. Eq (60) of \cite{cuff2013distributed}), when $R + 2\epsilon \geq I_{\bar{P}}(X;U|Z)+\gamma +\epsilon,$ we have
\[
\E_{\cb} [\dtv (\bar{P}_{X^n|Z^n=z^n, K=m'_0}, \bm Q_{X^n|Z^n=z^n,K=m'_0}) ] < \epsilon_n.
\]
Since $z^n$ are drawn i.i.d, and by Eq (61) of \cite{cuff2013distributed},
\begin{align*}
\dtv (\bm P_{X^nZ^n}, \bm Q_{X^nZ^n}) \leq 
&\dtv (\bm P_{X^nZ^n}, \bar{P}_{X^nZ^n}) \\
&+ \dtv (\bar{P}_{X^nZ^n}, \bm Q_{X^nZ^n}) \leq \epsilon_n.
\end{align*}
Therefore,
\begin{align}
    \dtv (\bm P_{X^nY^n}, &\bm Q_{X^nY^n}) \notag\\
    &\overset{\text{\ref{prop:marginal}}}{\leq} \dtv (\bm P_{X^nY^nZ^nMK}, \bm Q_{X^nY^nZ^nMK}) \notag\\
    &\overset{\text{\ref{prop:conditional}}}{=}\dtv (\bm P_{X^nZ^n}, \bm Q_{X^nZ^n})
    \leq \epsilon_n. \label{eq:tv-pq}
\end{align}
%
%
%
%
Then, we have
\begin{align*}
    & \phantom{ {}= } \E_{\cb} [ | \dtv (\bm P_{X^n}, \bm P_{Y^n}) - \dtv (\bar{P}_{X^n}, \bar{P}_{Y^n}) | ] \\
    &= \E_{\cb} [| \dtv (\bar{P}_{X^n}, \bm P_{Y^n}) - \dtv (\bar{P}_{X^n}, \bar{P}_{Y^n}) |] \\
    &\overset{\text{\ref{prop:triangle}}}{\leq} \E_{\cb} [\dtv (P_{Y^n}, \bar{P}_{Y^n})] \\
    &\overset{\text{\ref{prop:triangle}}}{\leq}\E_{\cb} [ \dtv (P_{Y^n}, \bm Q_{Y^n})
     + \dtv (\bar{P}_{Y^n}, \bm Q_{Y^n})] \\
    & \overset{\text{\ref{prop:marginal}}}{\leq} \E_{\cb} [\dtv (P_{X^nY^n}, \bm Q_{X^nY^n}) \\
      &\qquad\qquad + \dtv (\bar{P}_{Y^nZ^n}, \bm Q_{Y^nZ^n}) ]  \overset{\text{\eqref{eq:tv-pq}, \eqref{eq:tv-barpq}}}{\leq} 2\epsilon_n.
\end{align*}
\begin{align*}
    \text{So}\qquad & \E_{\cb} [\dtv (\bm P_{X^n}, \bm P_{Y^n})] \\
    \leq\  &\E_{\cb} [ | \dtv (\bm P_{X^n}, \bm P_{Y^n}) - \dtv (\bar{P}_{X^n}, \bar{P}_{Y^n}) | ] \\
    & +\E_{\cb} [\dtv (\bar{P}_{X^n}, \bar{P}_{Y^n}) ]
    \leq \Pi +\epsilon+2\epsilon_n.
\end{align*}

Next, we show that the average distortion requirement is also satisfied. 
First, we have
\begin{align}
    &\E [\dtv (\bm P_{X^nY^n}, \bar{P}_{X^nY^n})] \leq \E [\dtv (\bm P_{X^nY^n}, \bm Q_{X^nY^n}) \notag\\
    &+ \dtv (\bar{P}_{X^nY^n}, \bm Q_{X^nY^n})] \overset{\eqref{eq:tv-pq}}{\leq} \dtv (\bar{P}_{X^nY^n}, \bm Q_{X^nY^n}) + \epsilon_n \notag\\
    &\overset{\eqref{eq:tv-barpq},\eqref{prop:conditional}}{\leq} 2\epsilon_n.\label{eq:tv-prod-xy}
\end{align}
Notice that 
the sequence $(X^n, Y^n)$ drawn i.i.d. according to $\bar{P}$ is almost surely distortion typical (see, e.g., \cite{cover1999elements} Lemma 10.5.1), i.e., 
\[
\E_{\bar{P}_{X^nY^n}} [D(X^n,Y^n)] \leq \Delta + \epsilon\qquad a.s.
\]
where we note that the distortion and total variation sequences are both bounded thus uniformly integrable.
According to Lemma 5 of \cite{yassaee2014achievability},
\[
\E_{\bm P} [D(X^n, Y^n)] \leq \Delta + \epsilon_n d_{max}.
\]

The above proof is for strong perception. In the case of empirical perception, it follows from the line of argument in Theorem 2 of \cite{cuff2010coordination} that common randomness is not necessary. So the inequality $R+R_0\geq I(Y;U|Z)$ can be removed.
\end{IEEEproof}

\section{Converse}\label{sec:converse}
We use the time mixing technique for the proof.

\begin{IEEEproof}
    We first focus on the strong perceptual constraint.
    Suppose the tuple $(R, R_0, \Delta, \Pi)$ is achievable, then, given $\epsilon>0,$ there exists a sequence of $(n, 2^{n(R+\epsilon)}, 2^{n(R_0+\epsilon)})$ codes with encoding decoding functions $(f_n, g_n),$ such that
    \begin{align*}
        \E_{P} [D(X^n, Y^n)] &\leq \Delta + \epsilon \\
        \dtv ({P}_{X^n}, {P}_{Y^n}) &\leq \Pi +\epsilon.
    \end{align*}
    For fixed $n,$ let $M$ and $K$ denote the message and the common randomness, i.e.,
    \begin{align*}
        M &= f_n(X^n, Z^n, K) \\
        Y^n &= g_n(M,Z^n,K) = g_n(f_n(X^n, Z^n, K), Z^n, K).
    \end{align*}
    Let $T\sim \mathcal{U}([n])$ a uniform random variable over $[n].$ Also, let $U:= (T, M, K).$ Notice that $K\perp X^n$ and $X^n\perp Y^n|U,Z^n.$ 
    The rest of the proof follows closely that of the converse part of Theorem 2 in \cite{wagner2022rate}.
    We have
    \begin{align}\label{eq:converse1}
        n(R+\epsilon) &\geq H(M) \notag\\
        &\geq H(M|Z^n, K) \notag\\
        &\geq I(X^n; M| Z^n, K) \notag\\
        &= I(X^n; M,K | Z^n) \notag\\
        &= \sum_{t=1}^n I(X_t; M,K | Z^n, X^{t-1}) \notag\\
        &= \sum_{t=1}^n I(X_t; M,K, X^{t-1}, Z^{t-1}, Z_{t+1}^n |Z_t) \notag\\
        &\geq \sum_{t=1}^n I(X_t; M,K |Z_t) \notag\\
        &= nI(X_T; M,K | Z_T,T) \notag\\
        &= nI(X_T; M,K,T | Z_T) \notag\\
        &= nI(X_T;U |Z_T).
    \end{align}
    By the assumption of Theorem~\ref{thm:converse} under strong perceptual constraint, $Y^n$ and $Z^n$ are jointly i.i.d. sequences, so
    \begin{align}\label{eq:converse2}
        n(R+R_0+\epsilon) &\geq H(M,K) \notag\\
        &\geq H(M,K|Z^n) \notag\\
        &\geq I(Y^n;M,K|Z^n) \notag\\
        &= \sum_{t=1}^n I(Y_t; M,K | Z^n, Y^{t-1}) \notag\\
        &= \sum_{t=1}^n I(Y_t; M, K, Y^{t-1}, Z^{t-1}, Z_{t+1}^n | Z_t) \notag\\
        &\geq \sum_{t=1}^n I(Y_t; M,K | Z_t) \notag\\
        &= nI(Y_T; M,K |Z_T, T) \notag\\
        &= nI(Y_T; M,K,T |Z_T) \notag\\
        &= nI(Y_T; U|Z_T).
    \end{align}
    It is straightforward to verify that $\E_{P} [D(X, Y)]\leq \Delta$ using Definition \ref{defn:distortion}. When taking $n\rightarrow \infty$, we also have $\limsup_{n\rightarrow\infty}\dtv (P_{X^n}, P_{Y^n})\leq \Pi.$

    In the case of empirical perceptual constraint, the common randomness $K$ is not needed as discussed in Sec.~\ref{sec:achievability}. So $H(M)=H(M,K).$ Without likelihood encoding, $(X^n, Y^n)$ can no longer be treated as i.i.d. sequences.
    Therefore, inequality \eqref{eq:converse2} reduces to \eqref{eq:converse1}. As alluded in Sec.~\ref{sec:setup}, removing the common randomness results in a change of rate regions.
    Since $X\perp Y|U,Z,$ taking $U=Y,$ we have
    $
    R \geq I(X;Y|Z).
    $
    Suppose the tuple $(R,0,\Delta, \Pi)$ is achievable, then for $\epsilon>0,$
    \[
    \E_{\cb} [\dtv (\hat{P}_{X^n}, \hat{P}_{Y^n})] \leq \Pi +\epsilon.
    \]
     We observe that (see Property 2 of \cite{cuff2010coordination})
    \[
    \E [\hat{P}_{X^n}] = P_{X_{T}}, \quad
    \E [\hat{P}_{Y^n}] = P_{Y_{T}}.
    \]
    By convexity,
    \[
    \dtv (P_{X_{T}},P_{Y_{T}}) \leq \E [ \dtv (\hat{P}_{X^n}, \hat{P}_{Y^n}) ]
    \leq \Pi+\epsilon.
    \]
\end{IEEEproof}

\section{Unconstrained distortion and perception}\label{sec:discussion}
The coding theorem for extending the conventional rate-distortion theorem to the conditional case is given by \cite{gray1972conditional}.
When $\Pi>f(\Delta),$ the degree of perceptual fidelity is unconstrained, and the rate-distortion-perception functions (bounds) $R^{(e)}(\Delta, \infty)$ and $R^{(s)}(\Delta, \infty)$ reduce to the conditional rate-distortion function as discussed in Sec.~\ref{sec:setup}.

When the amount of distortion is unlimited, $\Delta\rightarrow\infty,$ the problem amounts to reconstructing sequences with empirical and product distributions similar to the source. If we require the perceptual constraint $\Pi$ to be asymptotically small, this becomes a special case of the coordination theory between two nodes studied in \cite{cuff2010coordination}. Specifically, for empirical perceptual constraint, the problem of communicating probability distributions \cite{kramer2007communicating} gives a rate-perception function. Results related to empirical coordination can often be proved using strong typicality \cite{cuff2010coordination, kramer2007communicating}. For strong coordination, stochastic encoding is crucial \cite{cuff2010coordination,cuff2013distributed}, so common randomness is important.

\section{Conclusion}
We consider the three-way trade-off between the rate, distortion, and perception inspired by the recent developments in using generative models for lossy compression to obtain realistic reconstructions, where the perceptual fidelity is evaluated by the total variation distance between both empirical and production distributions. We present rate-distortion-perception rate regions when common randomness and common side information are available to the encoder and decoder. A question that will be tackled in future work is whether common correlated side information can be used to reduce the need for common randomness when strong perceptual constraints are imposed.

\IEEEtriggeratref{11}
%
%
%

%
\bibliographystyle{IEEEtran}
\bibliography{ref}
%

\end{document}